\documentclass[12pt]{article}

\pdfoutput=1

\usepackage{amsmath,amsfonts,amssymb}
\usepackage{graphicx}
\usepackage{fouriernc}
\usepackage[T1]{fontenc}
\usepackage{fullpage}
\usepackage{setspace}
\usepackage[range-units=single, number-unit-product=~]{siunitx}
\usepackage{booktabs}
\usepackage[title]{appendix}
\usepackage[small]{titlesec}
\usepackage{xr}
\usepackage[hidelinks]{hyperref}
\usepackage{doi}
\usepackage{filecontents}

\usepackage[backend=bibtex,citestyle=authoryear-comp,bibstyle=authoryear,url=false,isbn=false,maxcitenames=2,maxbibnames=9,giveninits=true,uniquename=init,doi=true,dashed=false,sortcites=false]{biblatex}
\DeclareFieldFormat[article,inbook,incollection,inproceedings,patent,thesis,unpublished]{title}{\myhref{#1\isdot}}
\renewbibmacro{in:}{\ifentrytype{article}{}{\printtext{\bibstring{in}\intitlepunct}}}

\renewbibmacro*{volume+number+eid}{\printfield{volume}\setunit*{\addnbspace}\printfield{number}\setunit{\addcomma\space}\printfield{eid}}
\DeclareFieldFormat[article]{volume}{#1}
\DeclareFieldFormat[article]{number}{\mkbibparens{#1}}
\DeclareFieldFormat[article,inbook,incollection,inproceedings,patent,thesis,unpublished]{pages}{#1}
\newcommand{\myhref}[1]{%
  \ifboolexpr{%
    test {\ifhyperref}
    and
    not test {\iftoggle{bbx:url}}
    and
    not test {\iftoggle{bbx:doi}}
  }
  {\href{\doiorurl}{#1}}
  {#1}%
}

\usepackage{array}
\newcommand{\PreserveBackslash}[1]{\let\temp=\\#1\let\\=\temp}
\newcolumntype{C}[1]{>{\PreserveBackslash\centering}p{#1}}
\newcolumntype{R}[1]{>{\PreserveBackslash\raggedleft}p{#1}}
\newcolumntype{L}[1]{>{\PreserveBackslash\raggedright}p{#1}}

\renewcommand{\vec}[1]{\boldsymbol{#1}}
\newcommand{\T}{^\mathrm{T}}
\newcommand{\mat}[1]{\mathbf{#1}}

\newcommand\blfootnote[1]{%
  \begingroup
  \renewcommand\thefootnote{}\footnote{#1}%
  \addtocounter{footnote}{-1}%
  \endgroup
}

\DeclareSIUnit\cpkm{cpkm}
\DeclareSIUnit\cps{cps}
\DeclareSIUnit\day{day}
\DeclareSIUnit\year{yr}

\title{\textbf{Vertical-slice ocean tomography \\ with seismic waves}}
\author{J\"orn Callies$^1$, Wenbo Wu$^{1,2}$, Shirui Peng$^1$, Zhongwen Zhan$^1$ \\
  \footnotesize{$^1$California Institute of Technology, Pasadena, CA, USA}\\
  \footnotesize{$^2$Wood Hole Oceanographic Institution, Woods Hole, MA, USA}}
\date{}

\begin{document}

\maketitle
\blfootnote{Corresponding author: J\"orn Callies, jcallies@caltech.edu}

\section*{Key points}

\begin{itemize}
  \item Seismic \textit{T}~waves at different frequencies sample different parts of the water column.
  \item Frequency-dependent travel time changes between repeating earthquakes constrain the depth-dependent temperature change between the events.
  \item These data reveal the vertical structure of temperature anomalies produced by equatorial waves, mesoscale eddies, and decadal warming.
\end{itemize}

\clearpage

\section*{Abstract}

Seismically generated sound waves that propagate through the ocean are used to infer temperature anomalies and their vertical structure in the deep East Indian Ocean. These \textit{T}~waves are generated by earthquakes off Sumatra and received by hydrophone stations off Diego Garcia and Cape Leeuwin. Between repeating earthquakes, a \textit{T}~wave's travel time changes in response to temperature anomalies along the wave's path. What part of the water column the travel time is sensitive to depends on the frequency of the wave, so measuring travel time changes at a few low frequencies constrains the vertical structure of the inferred temperature anomalies. These measurements reveal anomalies due to equatorial waves, mesoscale eddies, and decadal warming trends. By providing direct constraints on basin-scale averages with dense sampling in time, these data complement previous point measurements that alias local and transient temperature anomalies.

\clearpage

\section*{Plain language summary}

Taking up almost all of the excess heat trapped on Earth by anthropogenic greenhouse gases, the ocean exerts a key control on our warming climate. Despite progress, tracking that heat remains an observational challenge. This study presents new measurements of ocean warming by making use of sound waves that are generated by earthquakes and propagate long distances through the ocean. These sound waves propagate faster in warmer seawater, so they arrive slightly earlier if warming has occurred. In this study, we measure such changes in arrival time at different frequencies---or pitches---that are sensitive to different parts of the water column, so warming in the upper ocean can be distinguished from warming in the deep ocean.

\clearpage

\section{Introduction}

The ocean is warming in response to accumulating greenhouse gases in the atmosphere, and its heat capacity dominates the climate system's thermal inertia. While the warming has been most pronounced in the surface ocean, heat transfer to the deep ocean importantly slows the climate change experienced at the surface \parencite[e.g.,][]{Hansen1984,Gregory2000,Held2010,Kostov2014}. \textcite{Roemmich2015}, for example, estimated from Argo floats that the top \SI{.5}{\kilo\meter} of the global ocean warmed at a rate of \SI{5}{\milli\kelvin\per\year} between 2006 and 2013, whereas the layer between \SIrange[range-phrase={ and }]{.5}{2}{\kilo\meter} warmed at a rate of \SI{2}{\milli\kelvin\per\year}. Had the heat not been transferred below \SI{.5}{\kilo\meter} depth, the top \SI{.5}{\kilo\meter} would have warmed more than twice as rapidly (excluding feedbacks with the atmosphere and radiative transfer). It is thus important to constrain the heat transfer to the deep ocean. Because this transfer is achieved by processes that depend on uncertain parameterizations in climate models (the overturning circulation, mesoscale eddies, and diapycnal mixing), strong observational constraints are crucial.

\begin{figure}[t]
  \centering
  \includegraphics[scale=0.85]{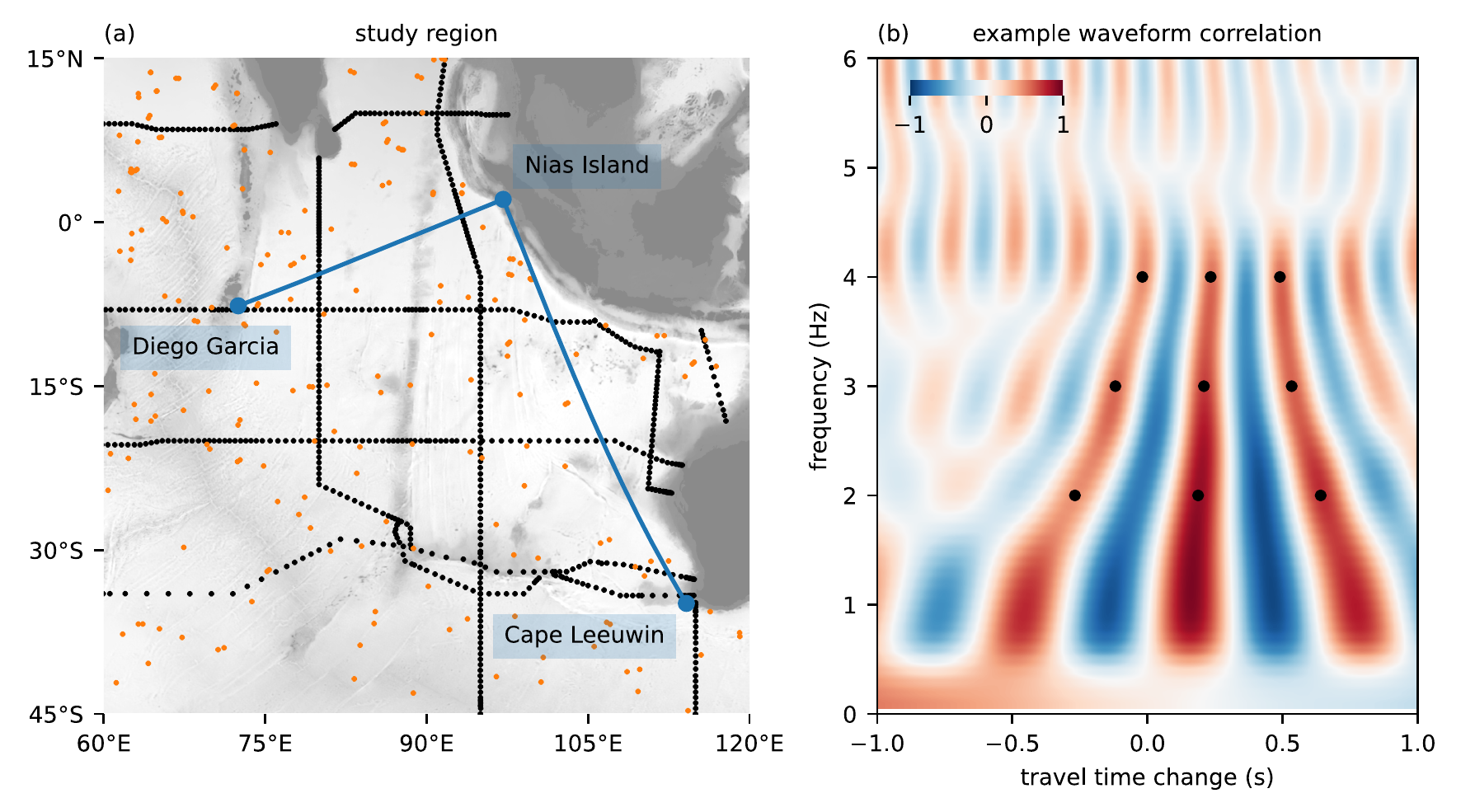}
  \caption{Study region and example measurement of the frequency-dependent travel time change. (a)~Locations of the Nias Island earthquakes and the hydrophone receivers at Diego Garcia and Cape Leeuwin. Also shown are the locations of all Argo profiles collected during the ten days following the 2005-03-28 Nias--Simeulue earthquake (orange dots) and the WOCE hydrographic sections (black dots). (b)~Frequency-dependent waveform correlation function for the event pair 2005-04-12 09:56:39, 2008-10-11 12:32:32. The black dots indicate measurements for the central maximum as well as the two adjacent maxima that are used for cycle skipping correction (not needed for this pair). An origin time correction of \SI{4.28}{\second} was applied based on the waveforms received at land station PSI.}
  \label{fig:map}
\end{figure}

It remains an observational challenge to isolate the small-amplitude and large-scale climate signal in the presence of much larger-amplitude but local fluctuations due to mesoscale eddies, internal waves, and other oceanic transients. To \SI{2}{\kilo\meter} depth and since the mid-2000s, Argo floats have provided unprecedented coverage of the world ocean. Even with currently about 4000~floats, however, the Argo array still aliases mesoscale eddies, and regional estimates of warming rates remain uncertain \parencite[e.g.,][Fig.~\ref{fig:map}a]{Wunsch2016,Dushaw2019}. Additionally, and maybe more glaringly, Core Argo floats do not sample half of the ocean's volume: that below \SI{2}{\kilo\meter} depth. While the Deep Argo array is expanding \parencite[e.g.,][]{Roemmich2019,Johnson2020d}, the extant record is short and limited to a few regions. Estimates of temperature change below \SI{2}{\kilo\meter} rely heavily on sparse hydrographic sections that are sampled about once a decade \parencite[e.g.,][Fig.~\ref{fig:map}a]{Roemmich1984,Purkey2010,Desbruyeres2016,Desbruyeres2017,Volkov2017}. State estimates constrain such changes by combining available observations with an ocean model \parencite[e.g.,][]{Wunsch2007}, but the possibility of model error and the lack of an uncertainty estimate make it difficult to assess how reliable such estimates are. For example, \textcite{Wunsch2014} inferred widespread cooling of the abyssal ocean using the ECCO (v4r1) state estimate (relative to an assumed and corrected initial state), whereas direct estimates from repeat hydrographic sections tend to indicate a dominance of warming \parencite[e.g.,][]{Purkey2010,Desbruyeres2017}. Clearly, better observational constraints are needed.

Recently, \textcite{Wu2020} demonstrated that sound waves generated by earthquakes, so-called \textit{T}~waves, can be used to constrain basin-scale temperature change in the deep ocean. These measurements are based on the idea of ``ocean acoustic tomography,'' which was originally proposed by \textcite{Munk1979} and successfully demonstrated at both planetary \parencite{Munk1989} and basin scales \parencite{ATOC1998}. \textit{T}~waves propagate faster in warmer water, so changes in the average temperature encountered by these waves can be detected as changes in their travel time. The use of sound waves produced by earthquakes eliminates the need to deploy synthetic sound sources. \citeauthor{Wu2020} used the seismic station DGAR on Diego Garcia to receive \textit{T}~waves generated \SI{2900}{\kilo\meter} away by earthquakes near Nias Island off Sumatra (Fig.~\ref{fig:map}a). To remove uncertainties in the source location and timing, repeating earthquakes were employed to extract the change in travel time between two events. A key advantage of such acoustic measurements over point measurements is that they intrinsically average the temperature change along the sound waves' path and therefore suffer much less from spatial aliasing.

In an accompanying manuscript (hereafter referred to as W23), we show that Comprehensive Nuclear-Test-Ban Treaty Organization (CTBTO) hydrophones are more sensitive \textit{T}-wave receivers than land stations like DGAR, which allows the detection of smaller earthquakes at the CTBTO station H08 off Diego Garcia than is possible with DGAR data. This use of CTBTO data improves the time resolution of the inferred deep-ocean temperature change between Nias Island and Diego Garcia. In W23, we further use Nias Island \textit{T}-waves received at the CTBTO station H01 off Cape Leeuwin to infer a time series of temperature change averaged along this \SI{4400}{\kilo\meter} section extending into the extra-tropical ocean (Fig.~\ref{fig:map}a).

\begin{figure}[t]
  \centering
  \includegraphics[scale=0.85]{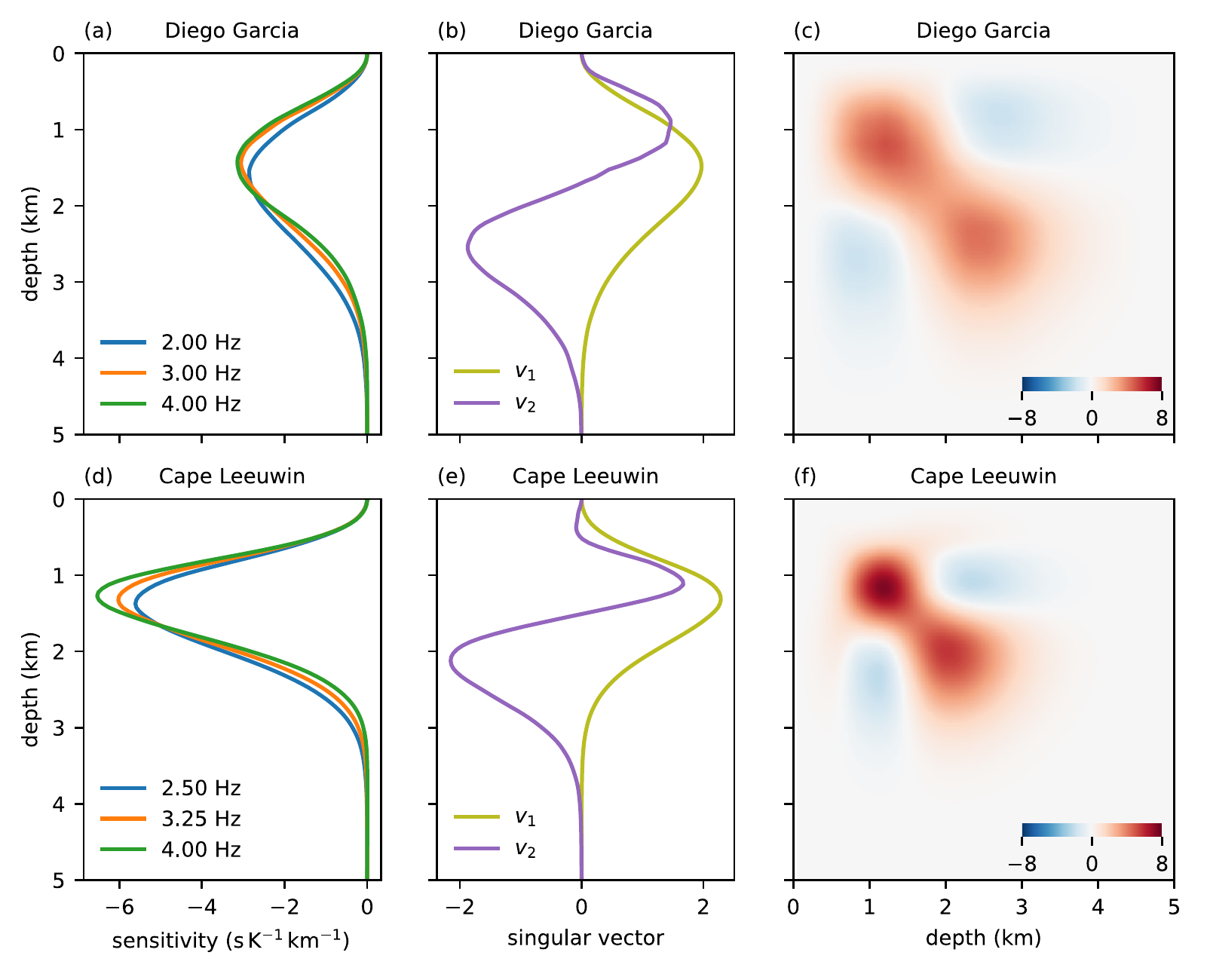}
  \caption{Sensitivity kernels and inference of vertical structure. Shown are (a),~(d)~the range-averaged sensitivity kernels at different frequencies, (b),~(e)~the first two singular vectors obtained from these range-averaged sensitivity kernels, and (c),~(f)~the resolution matrices of the SVD inversion employing the first two singular vectors. All these are shown for the path to Diego Garcia (top row) and that to Cape Leeuwin (bottom row).}
  \label{fig:kernels}
\end{figure}

Here, we make use of an additional advantage of the crisp arrivals of \textit{T}~waves in CTBTO hydrophone records: that changes in the travel time can be measured reliably at a number of different frequencies (Fig.~\ref{fig:map}b, Section~\ref{sec:inferring}). Because \textit{T}~waves at different frequencies are sensitive to different parts of the water column, this frequency dependence in the travel time change can be used to constrain the vertical structure of the temperature change \parencite[Fig.~\ref{fig:kernels}; cf.,][]{Munk1983,Shang1989}. This approach is similar to that employed in remote sensing of the atmosphere \parencite[e.g.,][]{Fu2004}, except that acoustic rather than electromagnetic waves are used.

A similar vertical-slice tomography scheme has been developed for acoustic measurements with synthetic sources \parencite[e.g.,][]{Munk1979,Munk1982b,Munk1995}, but it has not been realized at a basin scale. The sources employed in the Acoustic Thermometry of Ocean Climate (ATOC) experiment had frequencies centered on \SI{75}{\hertz} and were effectively point sources at precisely known locations, allowing for a convenient eigenray description of the propagation from source to receiver. These rays have a known geometry and, under certain circumstances, can be resolved and identified in arrival patterns \parencite{Spiesberger1980,OceanTomography1982,Cornuelle1993,ATOC1998,Worcester1999}. Relative changes in the arrival times of different rays can then be used to infer the vertical structure of temperature change, most easily in the range average. For example, the arrival time of steep rays that sample the surface mixed layer undergo a larger seasonal cycle than near-axial rays that are confined to the thermocline and deep ocean. On the bottom-mounted receivers used in ATOC, however, only relatively steep rays could be resolved and identified, and the information on the vertical structure of the temperature field was limited \parencite{Dushaw1999b,Dushaw2009a}.

The propagation and arrival patterns of \textit{T}~waves are more complicated \parencite[e.g.,][]{Okal2008}, but information on the vertical structure of the temperature change can still be extracted. Sensitivity kernels quantify how the arrival time changes in response to a temperature perturbation anywhere along the path and anywhere in the water column \parencite[Fig.~\ref{fig:kernels};][]{Wu2020}. Consistent with expectations from simplified modal calculations (Fig.~\ref{fig:modekernels}), these range-averaged sensitivity kernels shift upward in the water column for increasing frequencies (Fig.~\ref{fig:kernels}). A surface-intensified warming, for example, produces a larger reduction in travel time at high frequencies than at low frequencies. We use this frequency dependence to estimate a rough vertical structure of the range-averaged deep temperature anomalies. We find that the inferred structure of the anomalies along the two sections matches expectations based on the dynamics that produce the anomalies and is in general agreement with previous estimates, although the \textit{T}-wave data tends to show stronger anomalies and more reliably captures perturbations due to mesoscale eddies (Section~\ref{sec:timeseries}).

\section{Inferring vertical structure}
\label{sec:inferring}

The starting point for the inference of vertical structure in the temperature anomalies are the time series of travel time anomalies at a few different low frequencies, constructed from a total of \num{11479} repeating pairs that arise from \num{3457} earthquakes in 2005 to 2018 (Fig.~\ref{fig:map}b; W23). We choose \SIlist{2.00;3.00;4.00}{\hertz} for the path to Diego Garcia and \SIlist{2.50;3.25;4.00}{\hertz} for the path to Cape Leeuwin. A slightly higher minimum frequency is used for the latter path because \SI{2.00}{\hertz} \textit{T}~waves are less consistently received at Cape Leeuwin than at Diego Garcia. Measurements at higher frequencies are not reliable because the waveform correlation drops markedly beyond \SI{4}{\hertz} (e.g., Fig.~\ref{fig:map}b). For each frequency, we apply a Gaussian filter with width \SI{0.5}{\hertz} centered on that frequency before calculating the correlation function between the \textit{T}-wave arrivals of an event pair, as described in W23. How these time series are obtained from measured travel time differences between repeating events is described in \ref{sec:inversion}, and the cycle skipping corrections that are applied to the measurements are described in \ref{sec:csc}.

To turn the time series of travel time anomalies at different frequencies into an estimate of the evolving vertical structure of temperature anomalies, we perform a singular value decomposition (SVD) of the range-integrated sensitivity kernels (Fig.~\ref{fig:kernels}). The problem is severely under-determined, so we can only hope to obtain a coarse estimate of the vertical temperature structure. Let $\mat{K}$ denote the matrix whose three rows contain the range-integrated sensitivity kernels at the three frequencies, discretized to a $\Delta z = \SI{50}{\meter}$ grid. Then, at each event time~$t$, we would like to invert $\vec{\tau}(t) = \mat{K} \vec{T}(t) \Delta z$ for $\vec{T}(t)$, where $\vec{\tau}(t)$ contains the \textit{T}-wave travel time anomalies at the three frequencies, and $\vec{T}(t)$~contains the range-averaged temperature anomaly profiles on the same grid as the kernels. The SVD $\mat{K} = \mat{U} \vec{\Lambda} \mat{V}\T$ yields the singular values $\lambda_1 = \SI{2.6}{\second\per\kelvin\per\kilo\meter}$, $\lambda_2 = \SI{.21}{\second\per\kelvin\per\kilo\meter}$, and $\lambda_3 = \SI{0.022}{\second\per\kelvin\per\kilo\meter}$ for the path to Diego Garcia and $\lambda_1 = \SI{4.6}{\second\per\kelvin\per\kilo\meter}$, $\lambda_2 = \SI{.30}{\second\per\kelvin\per\kilo\meter}$, and $\lambda_3 = \SI{.012}{\second\per\kelvin\per\kilo\meter}$ for the path to Cape Leeuwin (see \ref{sec:inversion} for details). The rapid decay in the singular values is a result of the similarity of the sensitivity kernels at the chosen frequencies, i.e., their being nearly linearly dependent. Small singular values amplify errors (we only know the estimate~$\tilde{\vec{\tau}}$, not the true~$\vec{\tau}$), so a common trade-off must be made between resolution and precision. We choose to retain the first two singular vectors to obtain coarse vertical resolution with acceptable uncertainty: $\tilde{\vec{T}}_2(t) = h^{-1} \mat{V}_2 \vec{\Lambda}_2{}\!^{-1} \mat{U}_2{}\!\T \tilde{\vec{\tau}}(t)$, where $\mat{U}_2$ and $\mat{V}_2$ consist of the first two columns of $\mat{U}$ and $\mat{V}$, respectively, $\vec{\Lambda}_2$ is the diagonal matrix containing $\lambda_1$ and $\lambda_2$, and $h = \SI{5}{\kilo\meter}$ is a fixed reference depth. The estimate~$\tilde{\vec{T}}_2(t)$ is then related to the true temperature field~$\vec{T}(t)$ by $\tilde{\vec{T}}_2(t) = h^{-1} \mat{V}_2 \mat{V}_2{}\!\T \vec{T}(t) \Delta z + h^{-1} \mat{V}_2 \vec{\Lambda}_2{}\!^{-1}\mat{U}_2{}\!\T [\tilde{\vec{\tau}}(t) - \vec{\tau}(t)]$. In the absence of errors in $\tilde{\vec{\tau}}(t)$, the estimate $\tilde{\vec{T}}_2(t)$ is a projection of the true state $\vec{T}(t)$ onto the first two singular vectors, and the resolution matrix $\mat{V}_2 \mat{V}_2{}\!\T$ determines to what degree features can be resolved by the available data. For both paths, only features between about \SIrange[range-phrase={ and }]{.5}{3}{\kilo\meter} depth can be captured, and the depth resolution is no better than about \SI{1}{\kilo\meter} (Fig.~\ref{fig:kernels}). The projection coefficients are estimated from the data as $\tilde{\vec{c}}_2(t) = h^{-1} \mat{V}_2{}\!\T \tilde{\vec{T}}_2(t) \Delta z = h^{-1} \vec{\Lambda}_2{}\!^{-1} \mat{U}_2{}\!\T \tilde{\vec{\tau}}(t)$, with which the reconstructed temperature profile is the linear combination $\tilde{\vec{T}}_2(t) = \mat{V}_2 \tilde{\vec{c}}_2(t)$. (Note that the prior covariances for inferring $\tilde{\vec{\tau}}(t)$ from the travel time changes between repeating events are chosen such that the three components of $\tilde{\vec{c}}(t) = h^{-1} \vec{\Lambda}^{-1} \mat{U}\T \tilde{\vec{\tau}}(t)$ are uncorrelated. Any phase relations between them therefore arise entirely from the data. Details are given in \ref{sec:inversion}.)

Taking this SVD approach to inversion, we feign complete ignorance about the vertical structure of the range-averaged temperature anomalies. The inversion uses neither constraints from the physics that govern the temperature field nor prior information, say from previous measurements or model simulations. In particular, the inversion knows nothing of the typically strong surface intensification of temperature anomalies, which entails that inverted temperature anomalies tend to reach too deep and that upper-ocean anomalies can produce spurious oppositely signed deep anomalies (Fig.~\ref{fig:h08depth},~\ref{fig:h01depth}). We nevertheless choose this agnostic approach to illustrate most simply what information is contained in the \textit{T}-wave data. Future work should combine this information from seismic data with prior information and other observations.

\section{Time series}
\label{sec:timeseries}

\begin{figure}[p]
  \centering
  \includegraphics[scale=0.85]{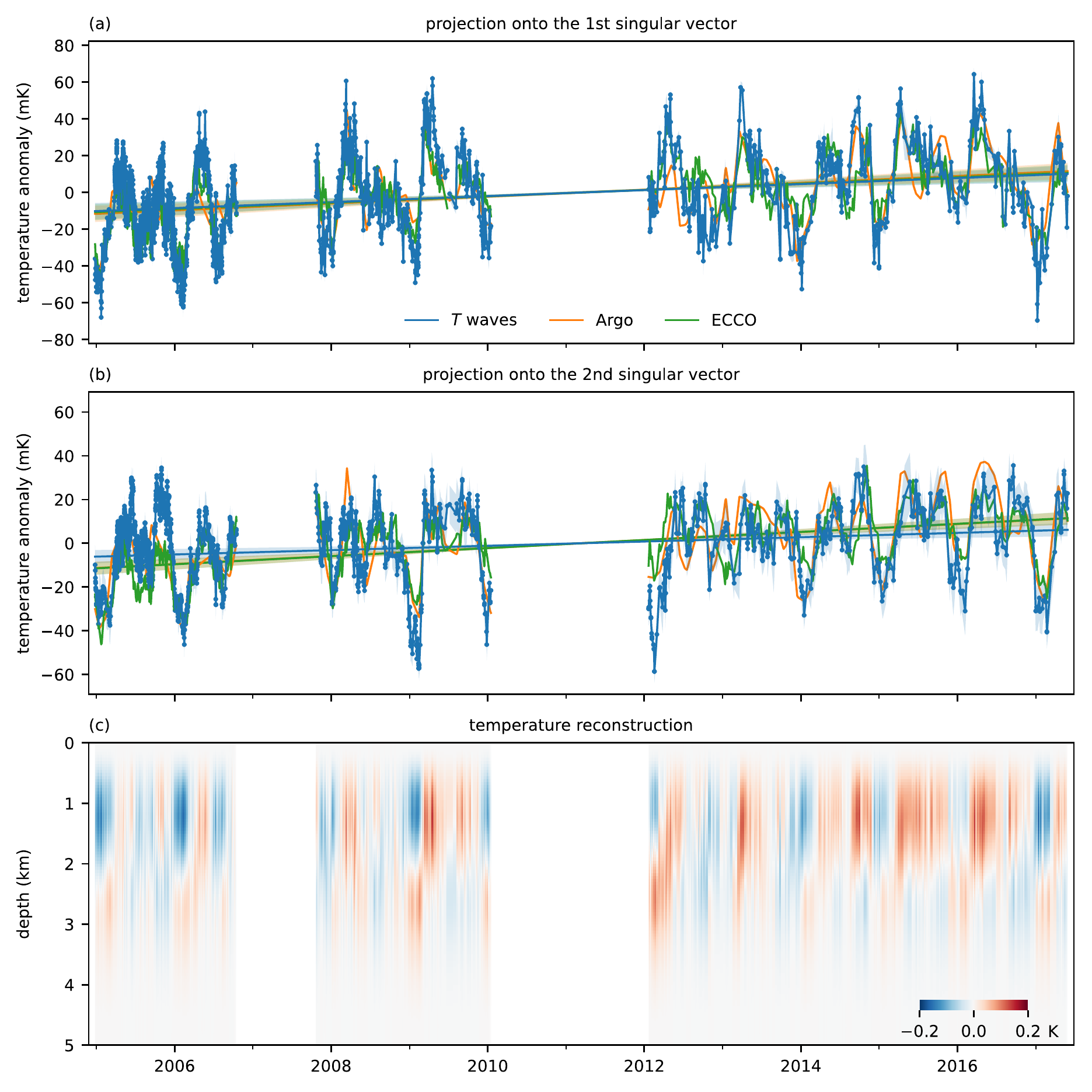}
  \caption{Inferred temperature anomalies for the path to Diego Garcia. Shown are (a),~(b)~the projections onto the first two singular vectors~$\tilde{\vec{c}}_2$ for the \textit{T}-wave data and two reference products and (c)~the SVD reconstruction of the range-averaged temperature anomaly profile~$\tilde{\vec{T}}_2$ obtained from the \textit{T}-wave data. Each dot in (a) and (b) corresponds to an estimate at an event time, and the shading indicates $\pm 2 \sigma$ confidence intervals around these estimates. Also shown are the estimated decadal trends of the projection coefficients~$\tilde{\vec{c}}_2$, including their $\pm 2 \sigma$ uncertainties.}
  \label{fig:h08}
\end{figure}

\begin{figure}[p]
  \centering
  \includegraphics[scale=0.85]{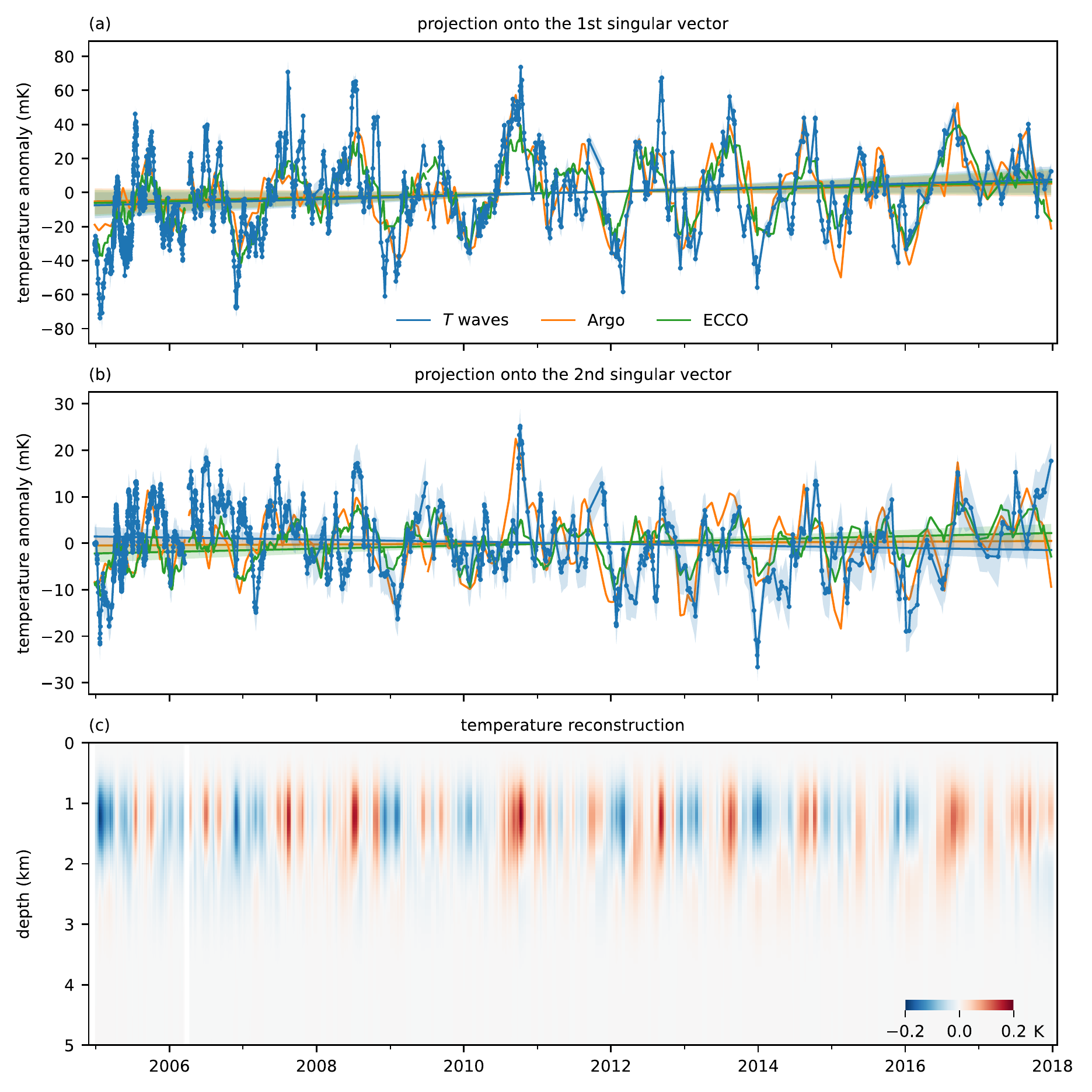}
  \caption{Inferred temperature anomalies for the path to Cape Leeuwin. Shown are (a),~(b)~the projections onto the first two singular vectors~$\tilde{\vec{c}}_2$ for the \textit{T}-wave data and two reference products and (c)~the SVD reconstruction of the range-averaged temperature anomaly profile~$\tilde{\vec{T}}_2$ obtained from the \textit{T}-wave data. Each dot in (a) and (b) corresponds to an estimate at an event time, and the shading indicates $\pm 2 \sigma$ confidence intervals around these estimates. Also shown are the estimated decadal trends of the projection coefficients~$\tilde{\vec{c}}_2$, including their $\pm 2 \sigma$ uncertainties.}
  \label{fig:h01}
\end{figure}

\begin{table}
  \centering
  \begin{tabular}{llcccccc}
    \toprule
    & & \multicolumn{2}{c}{trends (\si{\milli\kelvin\per\year})} & \multicolumn{2}{c}{12 mo.\ (\si{\milli\kelvin})} & \multicolumn{2}{c}{6~mo.\ (\si{\milli\kelvin})} \\
    & & 1 & 2 & 1 & 2 & 1 & 2 \\
    \midrule
    Diego Garcia & \textit{T}~waves & $+1.6 \pm 0.7$ & $+1.1 \pm 0.5$ & $15 \pm 4$ & $12 \pm 4$ & $15 \pm 4$ & $11 \pm 4$ \\
    & Argo & $+1.9 \pm 0.7$ & $+1.8 \pm 0.5$ & $\hphantom{0}9 \pm 4$ & $\hphantom{0}9 \pm 1$ & $12 \pm 4$ & $11 \pm 2$ \\
    & ECCO & $+1.7 \pm 0.7$ & $+1.9 \pm 0.5$ & $\hphantom{0}8 \pm 4$ & $\hphantom{0}8 \pm 2$ & $11 \pm 3$ & $11 \pm 3$ \\
    \midrule
    Cape Leeuwin & \textit{T}~waves & $+1.1 \pm 1.2$ & $-0.2 \pm 0.3$ & $20 \pm 5$ & $\hphantom{0}5 \pm 5$ & $\hphantom{0}3 \pm 5$ & $\hphantom{0}2 \pm 5$ \\
    & Argo & $+0.8 \pm 1.2$ & $+0.1 \pm 0.3$ & $19 \pm 5$ & $\hphantom{0}5 \pm 3$ & $\hphantom{0}3 \pm 4$ & $\hphantom{0}3 \pm 1$ \\
    & ECCO & $+1.0 \pm 1.2$ & $+0.3 \pm 0.3$ & $16 \pm 5$ & $\hphantom{0}3 \pm 3$ & $\hphantom{0}3 \pm 4$ & $\hphantom{0}3 \pm 2$ \\
    \bottomrule
  \end{tabular}
  \caption{Trends and amplitudes of the annual and semi-annual signals inferred from \textit{T}-wave data and previous products. These parameters are estimated for the temperature projections onto the first two singular vectors (columns labelled~1 and~2). All uncertainties are given as $\pm 2 \sigma$.}
  \label{tab:numbers}
\end{table}

Time series of the temperature profiles projected onto the first two singular vectors show prominent seasonal and sub-seasonal variations, as well as significant decadal trends for the path to Diego Garcia (Fig.~\ref{fig:h08}a,b;~\ref{fig:h01}a,b; Table~\ref{tab:numbers}). The first singular vectors have a similar shape as the kernels themselves, so their coefficients represent a weighted average of the deep temperature anomalies, with the weighting peaked at \SIlist{1.5;1.3}{\kilo\meter} for the paths to Diego Garcia and Cape Leeuwin, respectively (Fig.~\ref{fig:kernels}b,e). The second singular vectors have nodes at $\SIlist{1.7;1.5}{\kilo\meter}$ for the two paths, so the projection onto them is more sensitive to where in the water column the anomalies are located.

These projections inferred from \textit{T}-wave travel time anomalies show general agreement with previous products. We compare our estimates to monthly interpolated Argo data \parencite{Roemmich2009} and daily ECCO state estimate data \parencite[v4r4;][]{Forget2015,ECCO2021}. Using the sensitivity kernels, we infer travel time anomalies from the temperature anomalies of these products, interpolate those onto our event times, and subsequently treat them in the same way as the \textit{T}-wave anomalies.

For the path to Diego Garcia, seasonal and sub-seasonal variations in both these projections tend to line up (Fig.~\ref{fig:h08}a,b), indicating that the vertical structure inferred from the \textit{T}-wave data is broadly consistent with these previous estimates. There are notable exceptions. The anomalies inferred from the \textit{T}-wave data are stronger on average, particularly in the projection onto the first singular vector (Fig.~\ref{fig:h08}a, Table~\ref{tab:numbers}). The projection onto the second singular vector (Fig.~\ref{fig:h08}b) is more positive for the Argo product than for the \textit{T}-wave data in the latter part of the time series, and it is less negative for ECCO than for the \textit{T}-wave data in the early part of the time series, producing a stronger decadal trend in both previous products than inferred from the \textit{T}-wave data (Table~\ref{tab:numbers}).

For the path to Cape Leeuwin, the seasonal variations in both projections are similar between those inferred from \textit{T}-wave data and those from previous products (Fig.~\ref{fig:h01}a,b). In contrast to the equatorial path to Diego Garcia, the annual signal here is much stronger than the semi-annual signal (Table~\ref{tab:numbers}). The \textit{T}-wave data produces sizable spikes with a duration on the order of a month, which are typically missed by previous data (Fig.~\ref{fig:h01}a,b;~\ref{fig:h01depth};~\ref{fig:h01invdepth}). As discussed in W23, we interpret these spikes as resulting from mesoscale eddies traversing the path, most importantly those shed by the Leeuwin Current. The temperature anomalies of these eddies are largely confined to the thermocline \parencite[e.g.,][]{Fieux2005}, which is consistent with the anomalies in the two projections appearing in phase. At times, these eddies happen to be captured by Argo floats (e.g., 2005-10 and 2010-10), but they are more typically missed, as expected from the Argo float coverage (Fig.~\ref{fig:map}). ECCO does not capture mesoscale eddies because it has too low a horizontal resolution. The overall stronger variability along this path to Cape Leeuwin implies that decadal trends are more uncertain (Table~\ref{tab:numbers}), although the uncertainties in the estimation from the three data sources are not independent.

Time series of the temperature reconstruction $\tilde{\vec{T}}_2$ illustrate the information on the vertical structure contained in the \textit{T}-wave data. On the path to Diego Garcia, the inferred temperature anomalies are strongest in the upper \SI{2}{\kilo\meter} but reach substantially below this depth (Fig.~\ref{fig:h08}c). It cannot be inferred from the data whether these abyssal anomalies are real or whether they arise from the insufficient vertical resolution. (See Fig.~\ref{fig:h08depth},~\ref{fig:h01depth},~\ref{fig:h08invdepth}, and~\ref{fig:h01invdepth} for projections of Argo and ECCO data onto the first two singular vectors.)  Nevertheless, the anomalies display an upward phase propagation that is expected for long surface-generated equatorial waves that have a downward energy flux \parencite[e.g.,][]{Wunsch1977b,Philander1978,McPhaden1982,Luyten1982,Reppin1999}. This upward phase propagation is not apparent on the extratropical path to Cape Leeuwin (Fig.~\ref{fig:h01}c), where the inferred anomalies are more strongly confined to above \SI{2}{\kilo\meter} depth, consistent with mesoscale thermocline eddies.

In interpreting these results, it should be kept in mind that the sensitivity kernels are based on two-dimensional numerical simulations of the wave propagation from the source to the receiver \parencite{Wu2020}. These simulations depend, albeit not sensitively, on assumptions about the thickness and properties of sediment layers and on the neglect of off-geodesic effects. There is therefore representational uncertainty in our estimates of temperature anomalies and their vertical structure arising from uncertainty in the kernels. More work is needed to better understand these effects.

\section{Conclusions}

The data presented here demonstrate that changes in \textit{T}-wave travel times contain information on the vertical structure of the temperature anomalies encountered by these waves along their paths. This information can be extracted from travel time anomalies at a few low frequencies even though the sensitivity kernels at these frequencies have similar vertical structures (Fig.~\ref{fig:kernels}).

We here illustrated this vertical structure using a simple SVD inversion. In the future, the \textit{T}-wave data should be combined with Argo and ship-based hydrographic data, either using a relatively simple mapping as typically employed for Argo data or using state estimation as in ECCO, which also allows one to incorporate additional constraints, for example from altimetry and gravitometry. \textit{T} waves offer constraints on the large-scale temperature changes that are complementary to previous data. They intrinsically average in space, so they do not miss mesoscale eddies like the Argo array. They offer a dense sampling in time, which is important even for large-scale averages that still contain sizable anomalies induced by equatorial waves and mesoscale eddies. \textit{T}~waves offer constraints on the ocean below \SI{2}{\kilo\meter} depth, which has been sparsely sampled in space and time by ship-based hydrographic surveys. The vertical resolution obtained from \textit{T}-wave is relatively coarse, but they nevertheless constrain the vertical structure of the large-scale temperature anomalies.

In the present work, we restricted ourselves to low frequencies because the waveform correlation drops substantially at higher frequencies (Fig.~\ref{fig:map}b) and travel time changes cannot be extracted confidently. The \textit{T}-wave signals have plenty of power at these higher frequencies, so noise is unlikely to be the problem. Instead, we speculate that the higher-frequency signals contain multiple vertical acoustic modes. The modes become more confined in the vertical as the frequency increases, so more modes escape interaction with the bottom and subsequent attenuation. Different modes experience different time shifts, so the waveform correlation drops if multiple modes contribute substantially. If this interpretation is correct, much more detailed information on the vertical structure could be extracted if we were able to separate the modes in the received signal, for example using a vertical hydrophone array \parencite{DSpain2001}. Deploying such arrays is routine, so future \textit{T}-wave measurements could yield much stronger constraints on the vertical structure of the ocean's large-scale temperature variability.

\section*{Acknowledgements}

This material is based upon work supported by the Resnick Sustainability Institute and by the National Science Foundation under Grant No.~OCE-2023161.

\section*{Open research}

The IMS hydrophone data are available directly from the CTBTO upon request and signing a confidentiality agreement to access the virtual Data Exploitation Centre (vDEC). All seismic data were downloaded through the IRIS Data Management Center (\url{https://service.iris.edu/}), including the seismic networks II (GSN; \url{https://doi.org/10.7914/SN/II}), MY, PS, and GE (\url{https://doi.org/10.14470/TR560404}). The Global Seismographic Network (GSN) is a cooperative scientific facility operated jointly by the Incorporated Research Institutions for Seismology (IRIS), the United States Geological Survey (USGS) and the National Science Foundation (NSF), under Cooperative Agreement EAR-1261681. The processing code is available at \url{https://github.com/joernc/SOT}.

\section*{Disclaimer}

The views expressed in the paper are those of the authors and do not necessarily represent those of the CTBTO.

\printbibliography

\makeatletter\@input{xrsi.tex}\makeatother

\end{document}

\maketitle
\blfootnote{Corresponding author: J\"orn Callies, jcallies@caltech.edu}

\section*{Contents of this file}

\begin{enumerate}
  \item Texts S1 to S2
  \item Figures S1 to S9
  \item Table S1
\end{enumerate}

\clearpage

\renewcommand\thesection{Text S\arabic{section}}
\renewcommand\thefigure{S\arabic{figure}} 
\renewcommand\thetable{S\arabic{table}} 

\section{Inversion for travel time anomalies}
\label{sec:inversion}

The procedure to infer travel time anomalies relative to an unknown but common reference from measurements of \textit{T}-wave arrival time changes and origin time corrections (relative to a cataloged event time) is similar to that described in \textcite{Wu2020}, but here we improve on that inversion in a number of ways. In particular, we invert for the \textit{T}-wave delays and origin-time corrections separately, which allows us to more naturally use multiple \textit{T}-wave and land stations as well as a set of \textit{T}-wave frequencies. Instead of the somewhat \textit{ad hoc} regularization employed in \textcite{Wu2020}, we here impose a full set of prior and noise statistics and use a Gauss--Markov estimator. See, for example, \textcite{Wunsch2006} for an introduction to the methods employed here.

Our primary goal is to infer time series of the travel time anomalies at the observed frequencies~$\omega_1, \dots, \omega_l$, stacked into the vector
\begin{equation}
  \vec{\tau} = \begin{pmatrix}
    \begin{array}{c}
      \vec{\tau}_1 \\
      \vdots \\
      \vec{\tau}_l
    \end{array}
  \end{pmatrix}.
\end{equation}
These travel time anomalies, ascribed to changes in the ocean's temperature field, are differences between \textit{T}-wave arrival anomalies and origin time corrections:
\begin{equation}
  \vec{\tau} = \mat{D} \vec{a} \qquad \text{with} \qquad 
  \mat{D} = \begin{pmatrix}
    \begin{array}{cccc}
      \mat{I}_m & & & -\mat{I}_m \\
      & \ddots & & \vdots \\
      & & \mat{I}_m & -\mat{I}_m
    \end{array}
  \end{pmatrix} \qquad \text{and} \qquad
  \vec{a} = \begin{pmatrix}
    \begin{array}{c}
      \vec{a}^{(T)}_1 \\
      \vdots \\
      \vec{a}^{(T)}_l \\
      \vec{a}^{(L)}
    \end{array}
  \end{pmatrix},
\end{equation}
where \smash{$\vec{a}^{(T)}_i$} contains the \textit{T}-wave travel time anomalies for each event time at the different frequencies, and $\vec{a}^{(L)}$ contains the origin time corrections inferred from the land stations. Here, $\mat{I}_m$ denotes the identity matrix of size~$m$, the number of unique events. All delays are referenced to predicted arrival times based on the cataloged event times. The predictions are made using the PREM solid earth model for the land stations and a constant \textit{T}-wave reference velocity of \SI{1.5}{\kilo\meter\per\second}.

The travel time anomalies $\vec{\tau}$ are related to the range-averaged temperature anomalies through the sensitivity kernel:
\begin{equation}
  \vec{\tau} = (\mat{K} \otimes \mat{I}_m) \vec{T} \Delta z, \qquad \text{where} \qquad
  \mat{K} = \begin{pmatrix}
    \begin{array}{c}
      \mat{K}_1{}\!\T \\
      \vdots \\
      \mat{K}_l{}\!\T
    \end{array}
  \end{pmatrix}
  \qquad \text{and} \qquad
  \vec{T} = \begin{pmatrix}
    \begin{array}{c}
      \vec{T}_1 \\
      \vdots \\
      \vec{T}_n
    \end{array}
  \end{pmatrix}
\end{equation}
are the matrices containing the range-integrated kernels as rows and a stack of the time series of range-averaged temperature anomalies at the $n$~vertical levels, respectively, and~$\otimes$ denotes a Kronecker product. As described in the main text, we perform an SVD of the sensitivity matrix $\mat{K} = \mat{U} \vec{\Lambda} \mat{V}\T$. We normalize such that $\mat{U}\T \mat{U} = \mat{I}_3$ and $h^{-1} \mat{V}\T \mat{V} \Delta z = \mat{I}_l$, where $\Delta z = \SI{50}{\meter}$ is the vertical grid spacing and $h = \SI{5}{\kilo\meter}$ is a reference depth. With this normalization, the singular vectors are unitless and do not depend on the discretization (as long as it is fine enough). We denote the projections of the temperature anomalies onto the singular vectors by $\vec{c} = h^{-1} (\mat{V}\T \otimes \mat{I}_m) \vec{T} \Delta z$, so that $\vec{\tau} = h (\mat{U} \vec{\Lambda} \otimes \mat{I}_m) \vec{c}$, which can be inverted for the time series $\vec{c} = h^{-1} (\vec{\Lambda}^{-1} \mat{U}\T \otimes \mat{I}_m) \vec{\tau}$.

We cannot, however, observe the travel time anomalies directly. We only observe \emph{changes} in the travel time between repeating earthquakes. The design matrix thus takes differences of travel time anomalies between event pairs:
\begin{equation}
  \mat{E} = \begin{pmatrix}
    \begin{array}{cc}
      \mat{I}_l \otimes \mat{X}^{(T)} & \mat{0} \\
      \mat{0} & \mat{X}^{(L)}
    \end{array}
  \end{pmatrix}.
\end{equation}
The pair matrices $\mat{X}^{(T)}$ and $\mat{X}^{(L)}$ take these differences between events at all \textit{T}-wave and land stations involved. For example, if there were four pairs connecting five events observed at one \textit{T}-wave section, we might have
\begin{equation}
  \mat{X}^{(T)} = \begin{pmatrix}
    \begin{array}{ccccc}
      -1 & & & 1 & \\
      & -1 & 1 & \hphantom{-1} & \hphantom{-1} \\
      & -1 & & & 1 \\
      & & -1 & & 1
    \end{array}
  \end{pmatrix}.
\end{equation}
If pairs 1, 2, and 4 are observed at land station~1, and pairs~1 and~3 are observed on land station~2, the land station pair matrix would be
\begin{equation}
  \mat{X}^{(L)} = \begin{pmatrix}
    \begin{array}{ccccc}
      -1 & & & 1 & \\
      & -1 & 1 & \hphantom{-1} & \hphantom{-1} \\
      & & -1 & & 1 \\
      \hline
      -1 & & & 1 & \\
      & -1 &  &  & 1
    \end{array}
  \end{pmatrix}.
\end{equation}
The observed arrival time offsets, again measured relative to the cataloged event times, are then
\begin{equation}
  \vec{\delta} = \mat{E} \vec{a} + \vec{n} \qquad \text{with} \qquad
  \vec{\delta} = \begin{pmatrix}
    \begin{array}{c}
      \vec{\delta}^{(T)}_1 \\
      \vdots \\
      \vec{\delta}^{(T)}_l \\
      \vec{\delta}^{(L)}
    \end{array}
  \end{pmatrix}
\end{equation}
containing a stack of the \textit{T}-wave offsets measured at different frequencies and the land offsets. The observational noise is denoted by~$\vec{n}$.

We specify prior statistics for the projections onto the singular vectors (i.e., for~$\vec{c}$), and we assign these covariances to the \textit{T}-wave anomalies \smash{$\vec{a}^{(T)}_1, \dots, \vec{a}^{(T)}_l$}:
\begin{equation}
  \mat{R} = \begin{pmatrix}
    \begin{array}{cc}
      h^2 (\mat{U} \vec{\Lambda} \otimes \mat{I}_m) (\vec{\Sigma} \otimes \mat{C}) (\mat{U} \vec{\Lambda} \otimes \mat{I}_m)\T & \mat{0} \\
      \mat{0} & \mat{0}
    \end{array}
  \end{pmatrix} + \sigma^2 \mat{1}_{l+1} \otimes \mat{I}_m,
\end{equation}
where $\vec{\Sigma}$ is a diagonal matrix that specified the prior variances in the time series of the projections onto the singular vectors (Table~\ref{tab:priors}). The diagonal nature of $\vec{\Sigma}$ means that we assign no prior correlations between the projections onto the singular vectors. The matrix $\mat{C}$ specifies the time correlation $C_{ij} = e^{-|t_i-t_j|/\lambda}$, which we assume to be identical for all singular vectors. The expected standard deviation of origin time corrections is specified as $\sigma = \SI{5}{\second}$, and $\mat{1}_{l+1}$ denotes a square matrix of size $l+1$ filled with ones. This prescribes perfect correlation between the origin time correction and the part of the \textit{T}-wave anomalies that is due to origin time errors.

In addition to the random process with exponential correlation function, we also include deterministic annual and semi-annual cycles as well as a linear trend. We do so by appending their coefficients to the vector~$\vec{a}$ and extending the matrices $\mat{E}$, $\mat{R}$, and $\mat{D}$. For the design matrix,
\begin{equation}
  \mat{E} \to 
  \begin{pmatrix}
    \begin{array}{c|c}
      & \mat{A} \\
      \smash{\raisebox{.5\normalbaselineskip}{$\mat{E}$}} & \mat{0}
    \end{array}
  \end{pmatrix}
\end{equation}
with
\begin{equation}
  \mat{A} =
  \begin{pmatrix}
    \begin{array}{ccccc}
      \mat{I}_l \otimes \mat{X}^{(T)} \vec{t} & \mat{I}_l \otimes \mat{X}^{(T)} \cos \omega \vec{t} & \mat{I}_l \otimes \mat{X}^{(T)} \sin \omega \vec{t} & \mat{I}_l \otimes \mat{X}^{(T)} \cos 2\omega \vec{t} & \mat{I}_l \otimes \mat{X}^{(T)} \sin 2\omega \vec{t}
    \end{array}
  \end{pmatrix},
\end{equation}
where $\vec{t}$ contains the event times referenced to the midpoint between the first event and last event. For the prior covariance matrix,
\begin{equation}
  \mat{R} \to 
  \begin{pmatrix}
    \begin{array}{cc}
      \mat{R} & \mat{0} \\
      \mat{0} & h^2 (\mat{I}_5 \otimes \mat{U} \vec{\Lambda}) \vec{\Xi} (\mat{I}_5 \otimes \mat{U} \vec{\Lambda})\T
    \end{array}
  \end{pmatrix},
\end{equation}
where $\mat{\Xi}$ is a diagonal matrix containing the prior variances for the trend or seasonal amplitudes for the projections onto the singular vectors. Like for the random components, we assign no prior correlation between the trends and seasonal amplitudes of these projections. For the difference matrix,
\begin{equation}
  \mat{D} \to 
  \begin{pmatrix}
    \begin{array}{cc}
      \mat{D} & \mat{B}
    \end{array}
  \end{pmatrix},
\end{equation}
with
\begin{equation}
  \mat{B} =
  \begin{pmatrix}
    \begin{array}{ccccc}
      \mat{X}^{(T)} \vec{t} & \mat{X}^{(T)} \cos \omega \vec{t} & \mat{X}^{(T)} \sin \omega \vec{t} & \mat{X}^{(T)} \cos 2\omega \vec{t} & \mat{X}^{(T)} \sin 2\omega \vec{t}
    \end{array}
  \end{pmatrix},
\end{equation}
such that the trends and seasonal variations are included in the travel time anomalies~$\vec{\tau}$.

Once we specify the noise statistics $\mat{N} = \sigma_n{}\!^2 \mat{I}$ with $\sigma_n = \SI{0.01}{\second}$, we now have all information required for a Gauss--Markov estimate:
\begin{equation}
  \tilde{\vec{a}} = \mat{P} \mat{E}\T \mat{N}^{-1} \vec{\delta} \qquad \text{with} \qquad \mat{P} = \big( \mat{R}^{-1} + \mat{E}\T \mat{N}^{-1} \mat{E} \big)^{-1}.
\end{equation}
The desired estimate of the travel time anomalies is then $\tilde{\vec{\tau}} = \mat{D} \tilde{\vec{a}}$, and the posterior covariance~$\mat{P}$ can be propagated to this estimate as $\mat{D} \mat{P} \mat{D}\T$. Similarly, estimates of the projections onto singular vectors~$\tilde{\vec{c}}$, of the trends, and seasonal amplitudes of the signal can be obtained from~$\tilde{\vec{a}}$, and their uncertainties can be calculated from~$\mat{P}$.

The prescription $\sigma_n = \SI{.01}{\second}$ is meant to capture errors due to the changes in the source location between repeating events \parencite{Wu2020}, due to noise affecting the cross-correlation function, and due to hydrophone motion. The moored CTBTO hydrophones are displaced by local currents, which contributes measurement error because the moorings are not navigated. \textcite{Nichols2017} estimated that the CTBTO hydrophones at Wake Island, similar in design to the ones used here, can be displaced by 1/20 of the length of their mooring lines. The mooring lines for H08S2 and H01W2 are \SIlist{562;570}{\meter} long, respectively, producing maximum horizontal displacements of less than \SI{30}{\meter}. A horizontal displacement of \SI{30}{\meter} corresponds to a travel time anomaly of \SI{0.02}{\second}, which is equal to $2\sigma_n$. A more refined prescription of $\mat{N}$ might split the errors into their contributing factors and take the respective error correlations into account. We leave this refinement to future work.

\begin{table}
  \centering
  \begin{tabular}{l|c|S[table-format=2.0]S[table-format=2.0]S[table-format=2.0]|S[table-format=2.0]S[table-format=2.0]S[table-format=2.0]|S[table-format=2.0]S[table-format=2.0]S[table-format=2.0]|S[table-format=2.0]S[table-format=2.0]S[table-format=2.0]}
    \toprule
    & corr. time & \multicolumn{3}{c|}{random} & \multicolumn{3}{c|}{trend} & \multicolumn{3}{c|}{12\ mo.} & \multicolumn{3}{c}{6 mo.} \\
    & (days) & \multicolumn{3}{c|}{(\si{\milli\kelvin})} & \multicolumn{3}{c|}{(\si{\milli\kelvin\per\year})} & \multicolumn{3}{c|}{(\si{\milli\kelvin})} & \multicolumn{3}{c}{(\si{\milli\kelvin})} \\
    \midrule
    Diego Garcia & 15 & 15 & 10 & 5 & 4 & 2 & 1 & 15 & 10 & 5 & 15 & 10 & 5 \\
    Cape Leeuwin & 30 & 20 &  5 & 5 & 4 & 2 & 1 & 20 &  5 & 5 &  5 &  2 & 2 \\
    \bottomrule
  \end{tabular}
  \caption{Parameters of the prior covariances of the projections onto the singular vectors. Where three parameters are given, they are for the three singular vectors. For the seasonal signals, the cos and sin terms are each assigned half the indicated prior variance.}
  \label{tab:priors}
\end{table}

\section{Cycle-skipping correction}
\label{sec:csc}

Acoustic modes are dispersive. If the sounds speed profile is perturbed, a dispersive mode's phase and group will shift by different amounts. If the differential shift between a mode's phase and group is large enough, the correlation function will peak not at the correct phase shift but at the next peak over. The modal dispersion causes a cycle skip.

We here illustrate this process using a simple example of a Gaussian wave packet propagating through Munk's canonical sound speed profile. The observed \textit{T}-wave delays show a tell-tale sign of this process, suggesting that the same process occurs in the real system. The simple physics causing the cycle skipping allows for a robust identification and correction procedure, described below.

To illustrate the process causing cycle skipping, we consider modal propagation through a range-independent ocean with slowness profile~$S(z)$, following \textcite{Munk1995}. A signal propagating a distance~$L$ has a phase travel time $\tau = L s$, where $s = k/\omega$ is the modal phase slowness that equals the slowness~$S(z)$ at the turning depths. The modal structure $P$ is defined by
\begin{equation}
  \frac{\d^2 P}{\d z^2} + \left( \omega^2 S^2 - k^2 \right) P = 0
\end{equation}
with boundary conditions $P = 0$ at $z = 0$ (the surface) and $\d P / \d z = 0$ at $z = -h$ (the bottom). We apply the normalization $\int \d z \, P^2 = h$. The group slowness is
\begin{equation}
  s_\mathrm{g} = \frac{\d k}{\d \omega} = \frac{1}{s h} \int \d z \, S^2 P^2,
\end{equation}
where the integration is over the full depth. The mode's group travel time is $\tau_\mathrm{g} = L s_\mathrm{g}$.

We now perturb the slowness profile $S(z)$ to $S(z) + \Delta S(z)$. The difference in the modal phase slowness between the reference and perturbed profiles is
\begin{equation}
  \Delta s = \frac{1}{s h} \int \d z \, P^2 S \Delta S.
\end{equation}
The change in group slowness, in contrast, is
\begin{equation}
  \Delta s_\mathrm{g} = \frac{1}{s h} \int \d z \left[ \left( 2 - \frac{s_\mathrm{g}}{s} \right) P^2 + \omega (P^2)_\omega \right] S \Delta S,
\end{equation}
where $(P^2)_\omega = 2 P P_\omega$, and $P_\omega$ is defined by
\begin{equation}
  \frac{\d^2 P_\omega}{\d z^2} + \left( \omega^2 S^2 - k^2 \right) P_\omega = -2\left( \omega S^2 - k s_\mathrm{g} \right) P
\end{equation}
with the same boundary conditions as for $P$ and the additional constraint that $\int \d z \, P_\omega = 0$. Because variations in the slowness are much smaller than their mean values, the sensitivity profiles are roughly $P^2/h$ for the phase and $[P^2 + \omega (P^2)_\omega]/h$ for the group.

\begin{figure}[t]
  \centering
  \includegraphics[scale=0.85]{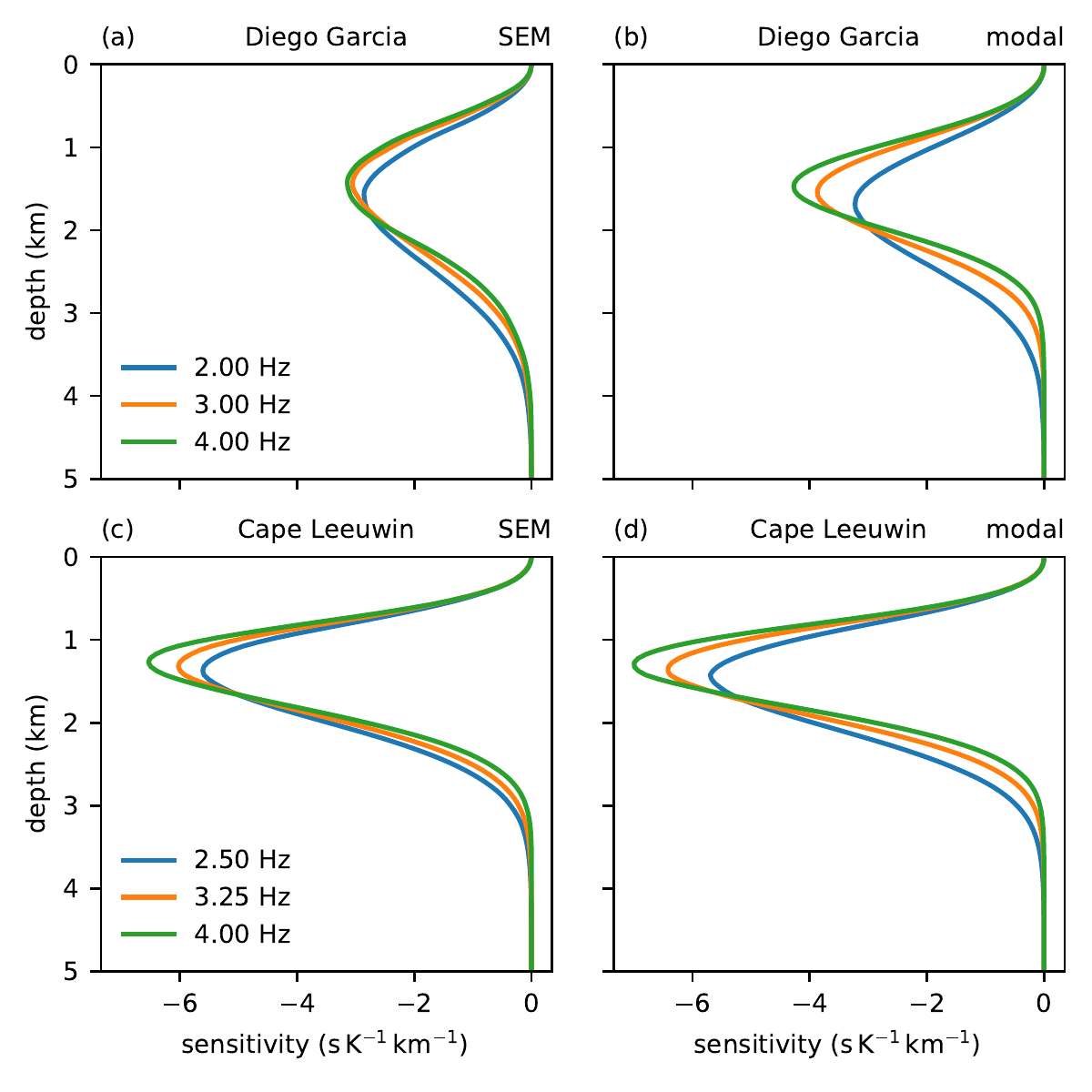}
  \caption{Sensitivity kernels calculated using two-dimensional numerical calculations as well as assuming the propagation of the first acoustic mode through a range-independent ocean. These kernels are shown for all considered frequencies and for both paths.}
  \label{fig:modekernels}
\end{figure}

\begin{figure}[ht]
  \centering
  \includegraphics[scale=0.85]{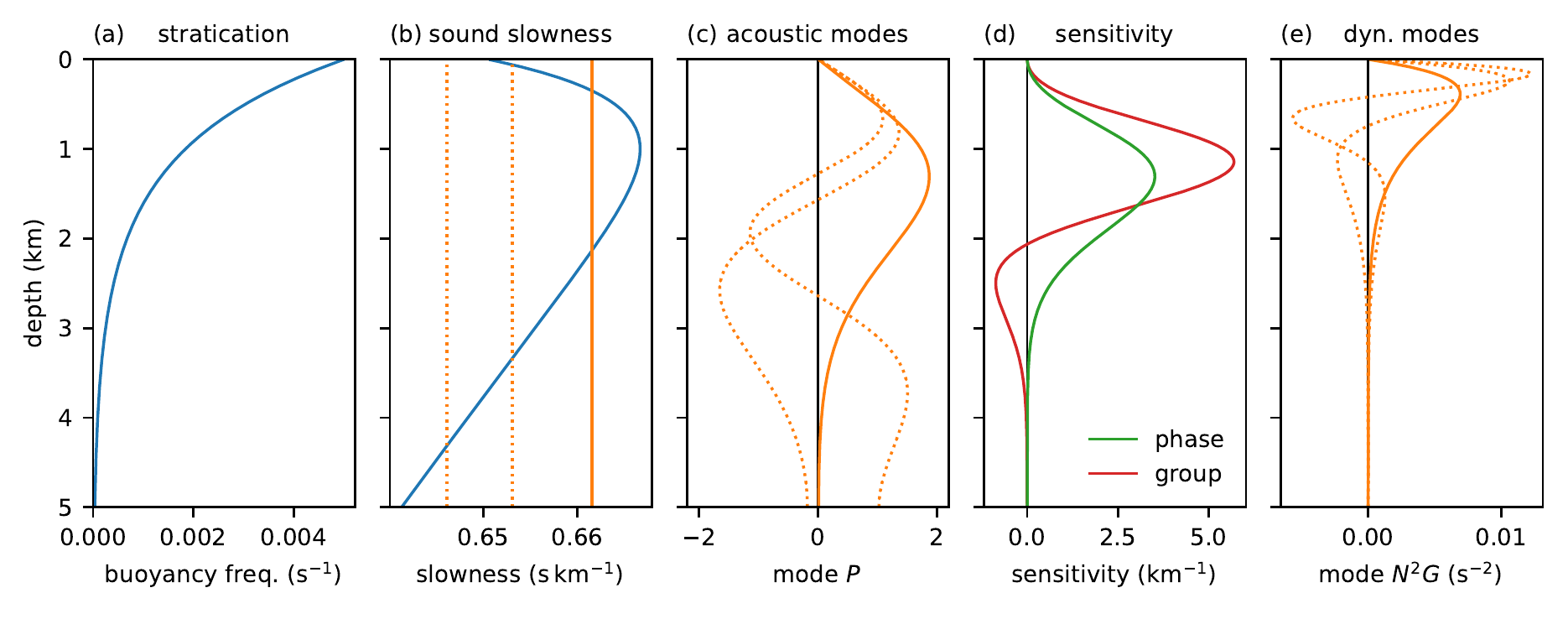}
  \caption{Illustration of dispersive propagation of low-frequency acoustic modes. Shown are profiles of (a)~the buoyancy frequency~$N$, (b)~Munk's canonical slowness profile~$S$ (blue) and the phase slownesses~$s$ of the first acoustic mode (solid orange) and two higher modes (dotted orange), (c)~the first acoustic mode (solid) and two higher modes (dotted), (d)~the phase and group sensitivities $P^2 S/sh$ (green) and $[(2-s_\mathrm{g}/s)P^2 + \omega (P^2)_\omega] S/sh$ (red) for the first acoustic mode, and (e)~the first dynamical mode (solid) and two higher modes (dotted).}
  \label{fig:dispersion}
\end{figure}

These phase and group sensitivity profiles thus have different structures (Fig.~\ref{fig:dispersion}). We illustrate their behavior using the lowest mode at $\omega/2\pi = \SI{2.5}{\hertz}$ for the canonical temperate profile of \textcite[][eq.~2.2.13]{Munk1974,Munk1995}. Above \SI{1.6}{\kilo\meter} depth, the group sensitivity to sound speed perturbations is larger than the phase sensitivity. If the temperature anomalies that cause these sound speed perturbations are dominated by this depth range, as is typical, the group delay $\Delta \tau_\mathrm{g} = L \Delta s_\mathrm{g}$ will exceed the phase delay $\Delta \tau = L \Delta s$. If the group delay is more than a half period larger than the phase delay, cycle skipping is likely. In this scenario, cycle skipping is always forward: the skip goes in the same direction as the delay. It is important to correct this cycle skipping because otherwise large anomalies would be overestimated systematically.

As an example, consider a slowness perturbation that is caused by a single low dynamical mode (Fig.~\ref{fig:dispersion}). For the buoyancy frequency profile $N = N_0 e^{z/d}$ underlying the canonical sound speed profile, the sound speed perturbation is proportional to~$N^2 G$. The dynamical mode~$G$ for vertical displacement is defined by
\begin{equation}
  \frac{\d^2 G}{\d z^2} + \frac{N^2}{c^2} G = 0
\end{equation}
with $G = 0$ at $z = 0$ and $z = -h$, and a normalization $\int \d z \, N^2 G^2 = h$ is applied. We use $h = \SI{5}{\kilo\meter}$, but this has little impact on the results because neither the first acoustic mode nor the first dynamical mode has much amplitude in the bottom \SI{1}{\kilo\meter} or so. For the first dynamical, $\Delta \tau_\mathrm{g} / \Delta \tau = 1.61$, so cycle skipping is likely for $|\Delta \tau| > \SI{.33}{\second}$. This is about where we observe cycle skipping to occur in the real \textit{T}~waveforms received at Cape Leeuwin.

\begin{figure}[t]
  \centering
  \includegraphics[width=.615\textwidth]{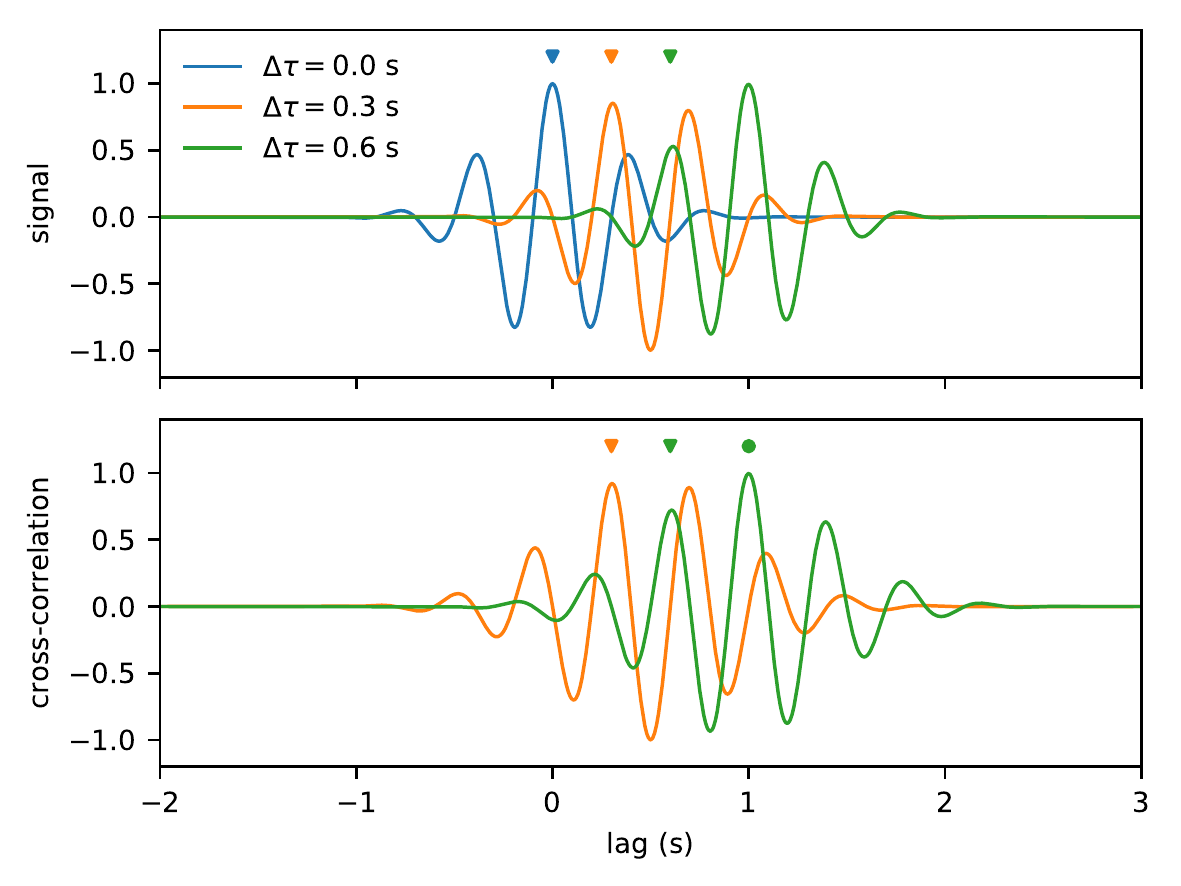}
  \caption{Illustration of cycle skipping using synthetic signals. The top panel shows three signals that have experienced different amounts of phase lags $\Delta \tau$ and group lags $\Delta \tau_\mathrm{g} = 1.61 \Delta \tau$. The triangles trace the peak that is at the center of the signal without lag, illustrating the phase shift. The bottom panel shows the cross-correlation functions between the lagged signals and the reference signal without lag. The maximum of the cross-correlation function occurs at the correct phase lag (marked by triangle) for the signal with $\Delta \tau = \SI{.3}{\second}$. Cycle skipping occurs for the signal with $\Delta \tau = \SI{.6}{\second}$: the maximum (marked by circle) is offset by one period from the correct lag, which coincides with a secondary maximum (marked by triangle)}
  \label{fig:synthcc}
\end{figure}

To further illustrate the cycle skipping, we consider a signal with the source function $e^{-\sigma^2 t^2} \cos \omega t$. We again use the frequency $\omega/2\pi = \SI{2.5}{\hertz}$, and we impose a bandwidth $\sigma/2\pi = \SI{.5}{\hertz}$ corresponding to the filtering we use for the real data. We then propagate this signal assuming a dispersion relation linearized around $\omega$. We impose phase lags $\Delta \tau = \SIlist{.0;.3;.6}{\second}$ and group lags $\Delta \tau_\mathrm{g} = 1.61 \Delta \tau$, corresponding to propagation through an anomaly caused by the first dynamical mode (Fig.~\ref{fig:synthcc}). Correlating the delayed signals with the signal without lag, the signal with $\Delta \tau = \SI{.3}{\second}$ produces a maximum of the correlation function at the correct phase lag. But the signal with $\Delta \tau = \SI{.6}{\second}$ experiences cycle skipping: the maximum of the correlation function shifts by one period to a lag of \SI{1.0}{\second}. The secondary maximum to the left is at the correct lag. If cycle skipping can be identified, it can easily be corrected by picking the adjacent peak in the correlation function.

\begin{figure}[ht]
  \centering
  \includegraphics[scale=0.85]{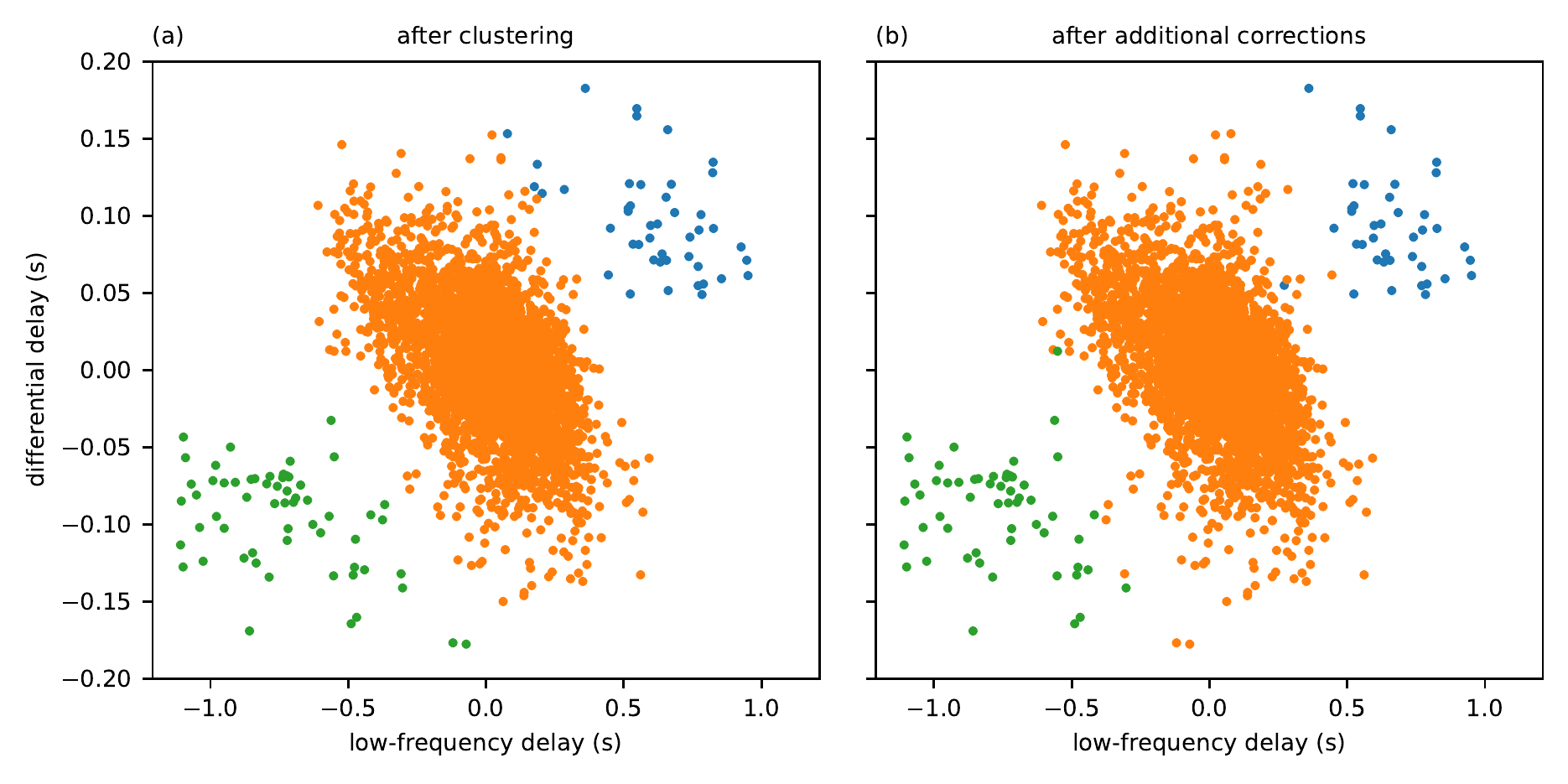}
  \caption{Cycle-skipping corrections for the path to Diego Garcia. Shown are the low-frequency delays (\SI{2.0}{\hertz}) vs.\ the differential delays (\SI{4.0}{\hertz} minus \SI{2.0}{\hertz}). The green dots indicate identified cycle skipping that is corrected to the right, the blue dots indicate cycle skipping corrected to the left, and orange dots indicate pairs not identified as being affected by cycle skipping. The identifications are (a)~after the cluster analysis and (b)~after addition corrections (or reversals) based on the inversion cost.}
  \label{fig:cluster-h08}
\end{figure}

\begin{figure}[ht]
  \centering
  \includegraphics[scale=0.85]{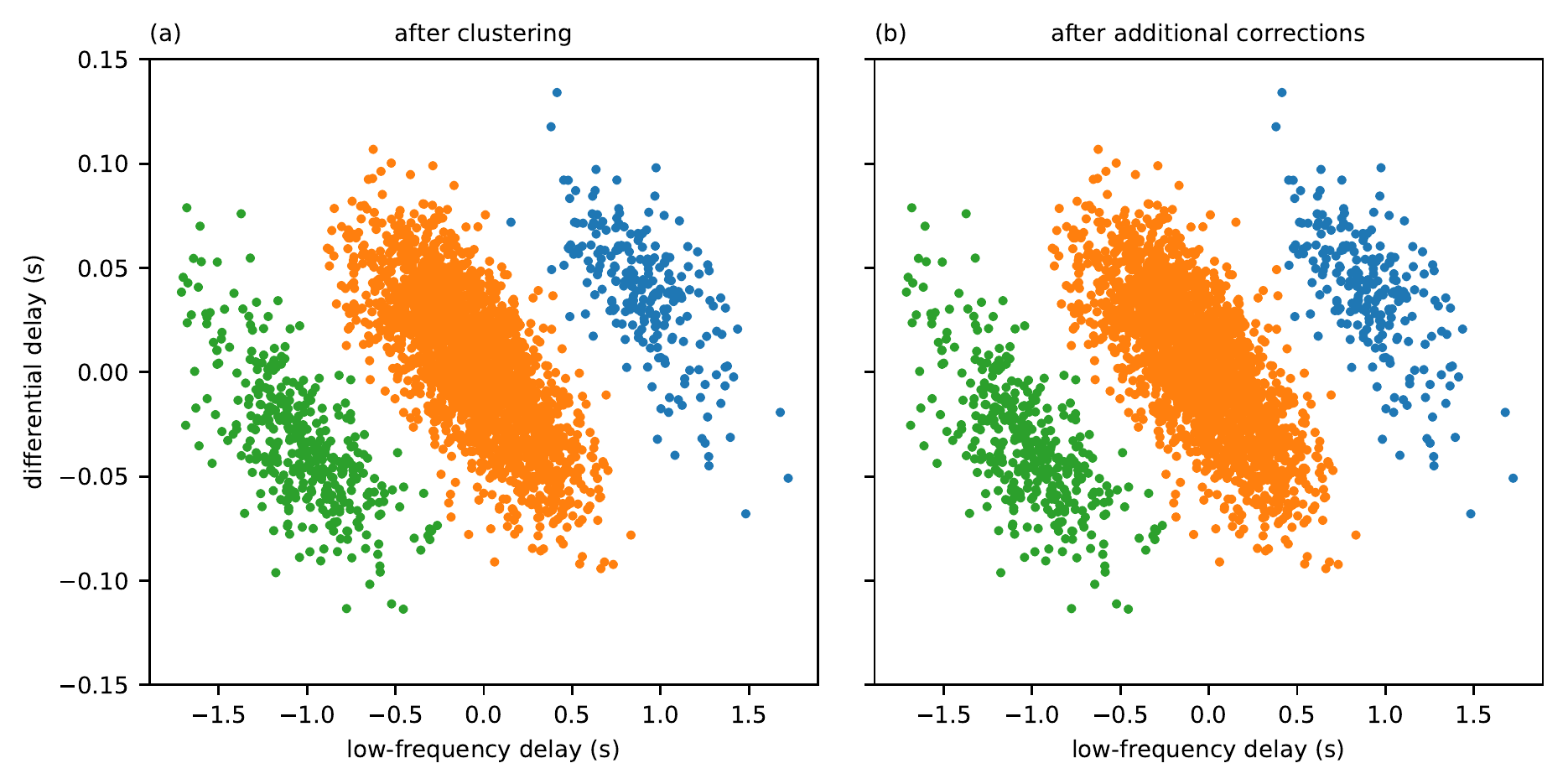}
  \caption{Cycle-skipping corrections for the path to Cape Leeuwin. Shown are the low-frequency delays (\SI{2.5}{\hertz}) vs.\ the differential delays (\SI{4.0}{\hertz} minus \SI{2.5}{\hertz}). The green dots indicate identified cycle skipping that is corrected to the right, the blue dots indicate cycle skipping corrected to the left, and orange dots indicate pairs not identified as being affected by cycle skipping. The identifications are (a)~after the cluster analysis and (b)~after addition corrections (or reversals) based on the inversion cost.}
  \label{fig:cluster-h01}
\end{figure}

This type of cycle skipping caused by dispersive modal propagation can be identified in the real data. If the first acoustic mode dominates and temperature anomalies are surface-intensified, the cycle skipping causes an increase in the magnitude of the low-frequency (say, \SI{2.5}{\hertz}) phase delay. At the same time, it causes a decrease in the differential delay between a higher frequency (say, \SI{4}{\hertz}) and this low frequency. Plotting the measured low-frequency and differential delays against one another reveals three distinct clusters: a center cluster with small low-frequency delays that is not affected by cycle skipping and adjacent clusters with larger-magnitude low-frequency delays that suffer from forward or backward cycle skipping (Fig.~\ref{fig:cluster-h08},~\ref{fig:cluster-h01}).

We identify pairs in these clusters by employing a Gaussian mixture model with three components that have shared covariances. We correct pairs in the left cluster by shifting to the next maximum of the correlation function to the right of the original maximum, and we correct pairs in the right cluster by shifting to the next maximum to the left. We then probe for additional cycle skips (or reversals of corrections applied after the cluster analysis) by looking for reductions in the cost of the inversion for travel time anomalies \parencite[cf.][]{Wu2020}. We allow such additional corrections if the Gaussian mixture model indicates a probability greater than 0.1\,\% to belong to a cluster different from that identified as most likely. For the path to Diego Garcia, this procedure corrects 40~pairs to the left and 56~pairs to the right (out of a total of 3831~used pairs). For the path to Cape Leeuwin, cycle skipping is more common, both because the path is longer and because travel time anomalies are somewhat larger. Here, 230~pairs are corrected to the left, and 393~pairs are corrected to the right (out of a total of 3032~used pairs).

\begin{figure}[p]
  \centering
  \includegraphics[scale=0.85]{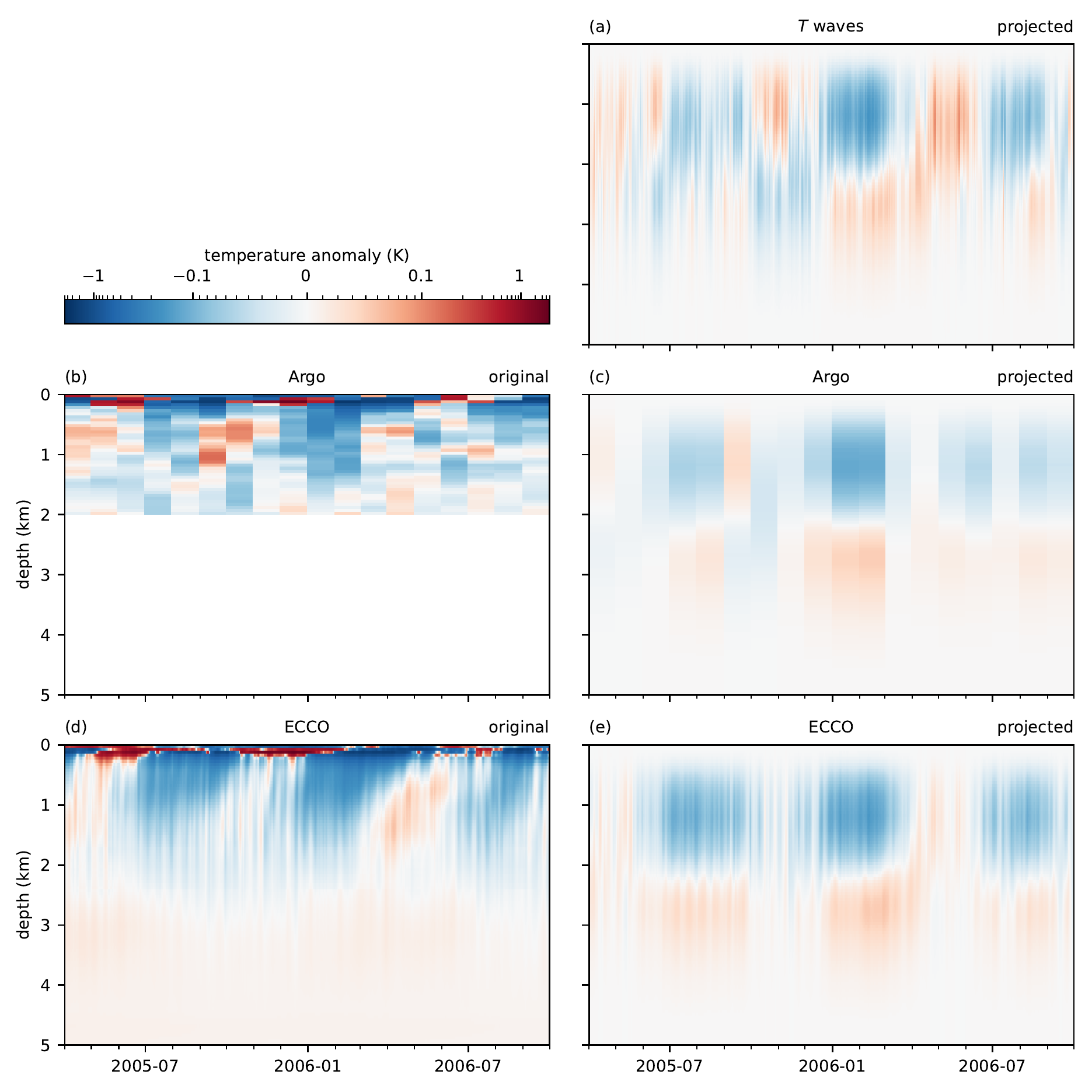}
  \caption{Temperature anomalies for the path to Diego Garcia for the early part of the observed period. Shown are (b),~(d)~the anomalies from the Argo and ECCO products and (a),~(c),~(e)~the projections onto the first two singular vectors inferred from \textit{T}~waves and applied to the Argo and ECCO products. Note the arsinh-scaled color map.}
  \label{fig:h08depth}
\end{figure}

\begin{figure}[p]
  \centering
  \includegraphics[scale=0.85]{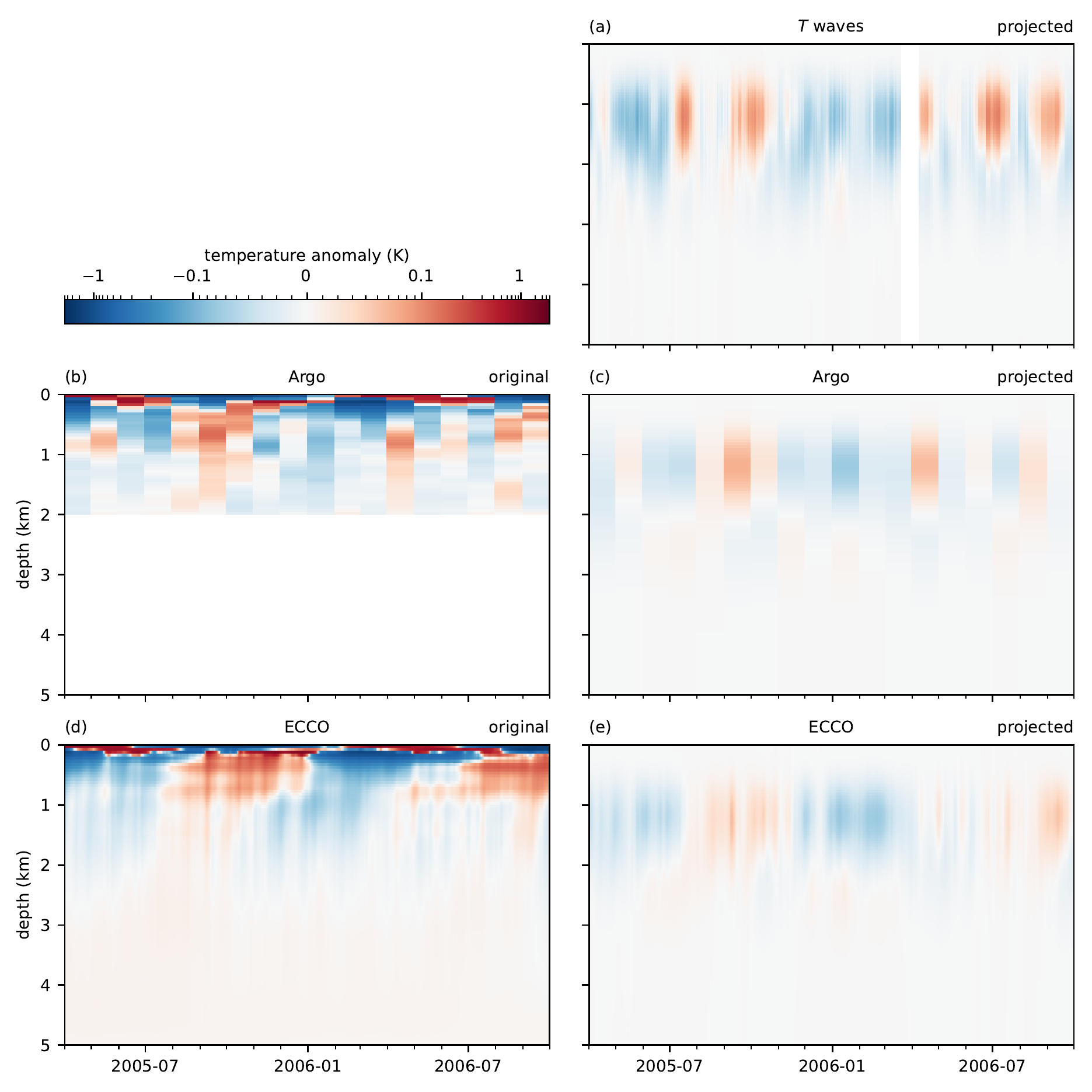}
  \caption{Temperature anomalies for the path to Cape Leeuwin for the early part of the observed period. Shown are (b),~(d)~the anomalies from the Argo and ECCO products and (a),~(c),~(e)~the projections onto the first two singular vectors inferred from \textit{T}~waves and applied to the Argo and ECCO products. Note the arsinh-scaled color map.}
  \label{fig:h01depth}
\end{figure}

\begin{figure}[p]
  \centering
  \includegraphics[scale=0.85]{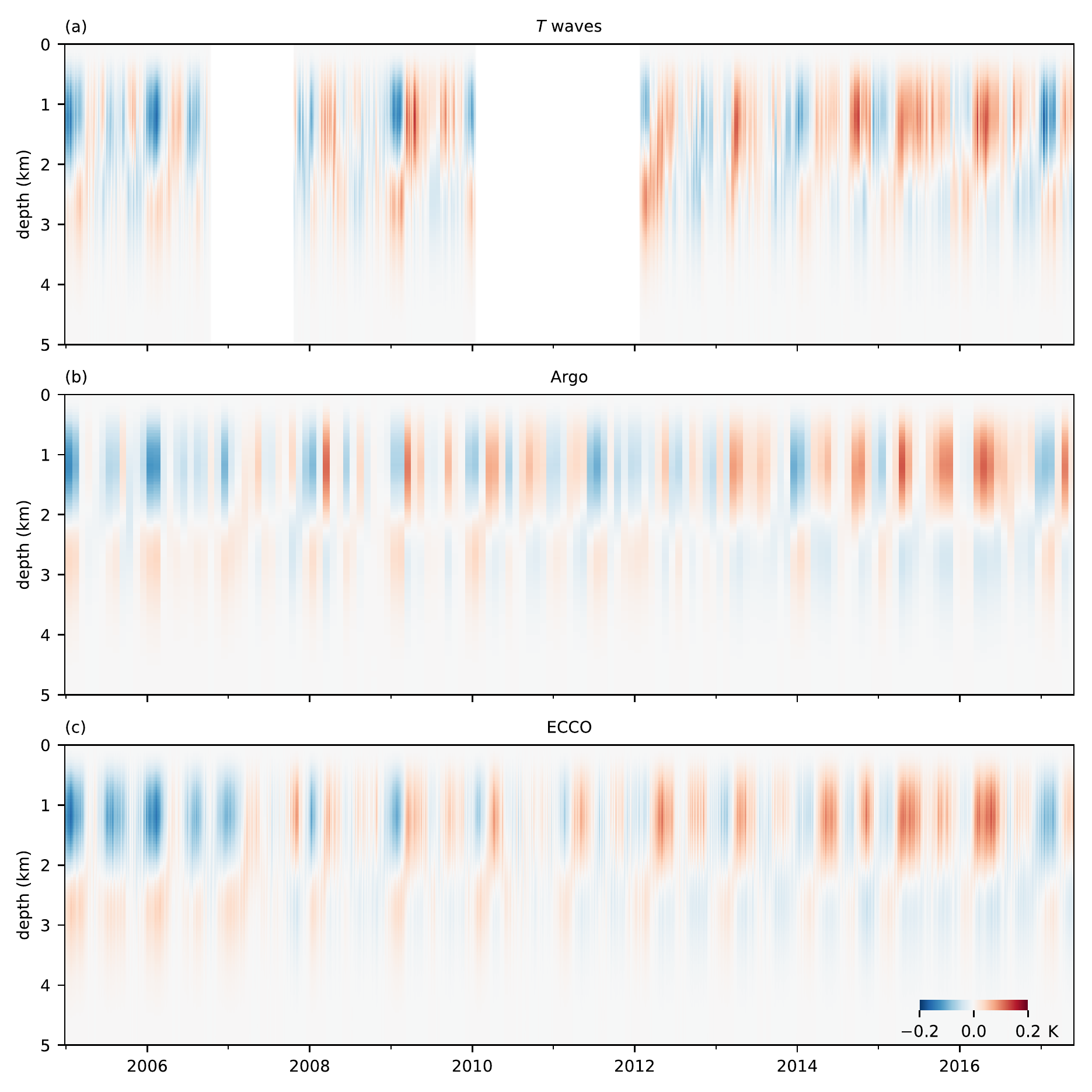}
  \caption{Temperature anomalies for the path to Diego Garcia projected onto the first two singular vectors. Shown are (a)~the time series inferred from \textit{T}-wave data, (b)~the projections of anomalies from the Argo product, and (c)~projections of anomalies from the ECCO product.}
  \label{fig:h08invdepth}
\end{figure}

\begin{figure}[p]
  \centering
  \includegraphics[scale=0.85]{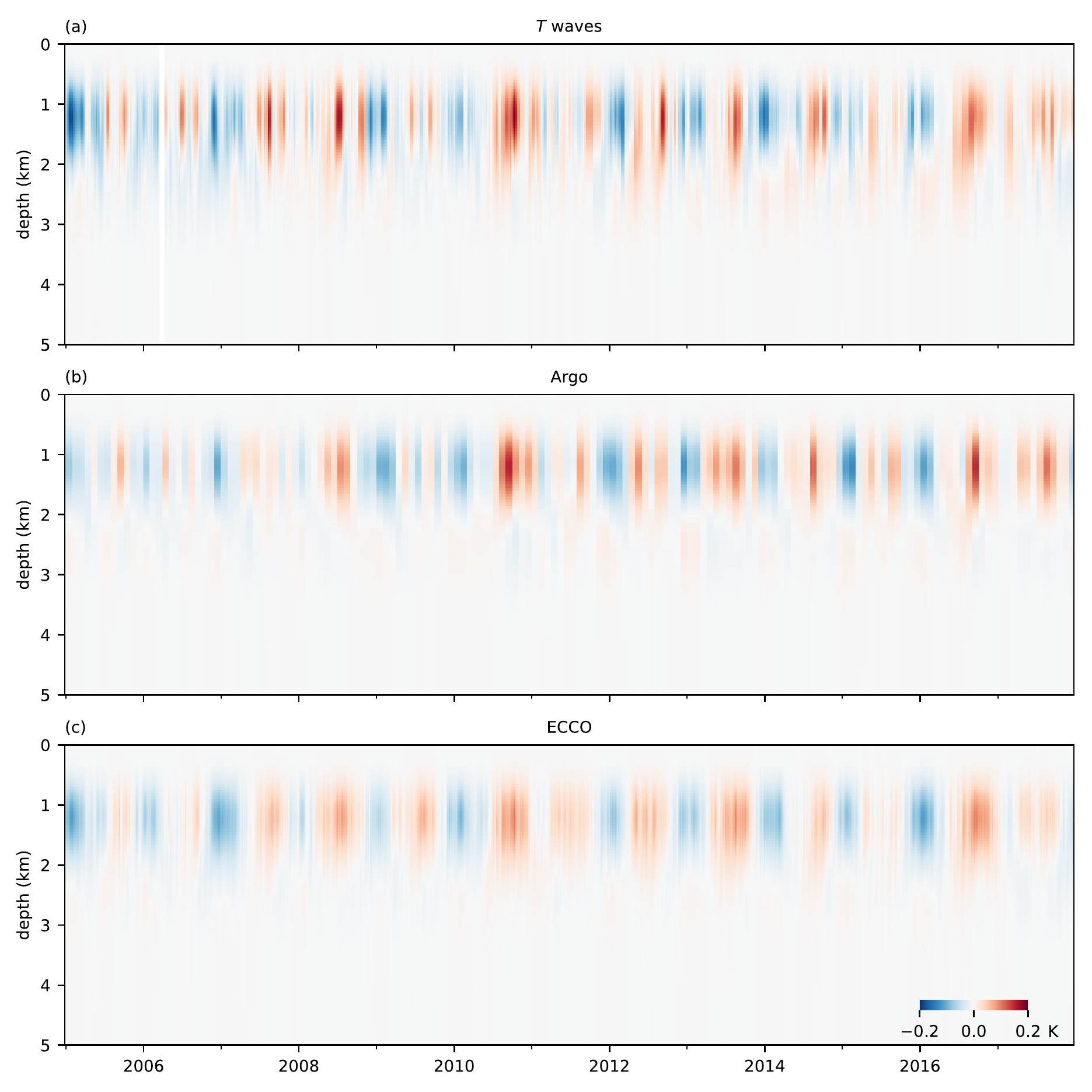}
  \caption{Temperature anomalies for the path to Cape Leeuwin projected onto the first two singular vectors. Shown are (a)~the time series inferred from \textit{T}-wave data, (b)~the projections of anomalies from the Argo product, and (c)~projections of anomalies from the ECCO product.}
  \label{fig:h01invdepth}
\end{figure}

\printbibliography

\makeatletter\@input{xrms.tex}\makeatother